# Low Radio Frequency Observations from the Moon Enabled by NASA Landed Payload Missions

Jack O. Burns[1,2], Robert MacDowall[3], Stuart Bale[4], Gregg Hallinan[5], Neil Bassett[2], Alex Hegedus[6]

Abstract

A new era of exploration of the low radio frequency Universe from the Moon will soon be underway with landed payload missions facilitated by NASA's Commercial Lunar Payload Services (CLPS) program. CLPS landers are scheduled to deliver two radio science experiments, ROLSES to the nearside and LuSEE to the farside, beginning in 2021. These instruments would be pathfinders for a 10-km diameter interferometric array, FARSIDE, composed of 128 pairs of dipole antennas proposed to be delivered to the lunar surface later in the decade. ROLSES and LuSEE, operating at frequencies from ≈100 kHz to a few tens of MHz, will investigate the plasma environment above the lunar surface and measure the fidelity of radio spectra on the surface. Both use electrically-short, spiral-tube deployable antennas and radio spectrometers based upon previous flight models. ROLSES will measure the photoelectron sheath density to better understand the charging of the lunar surface via photoionization and impacts from the solar wind, charged dust, and current anthropogenic radio frequency interference. LuSEE will measure the local magnetic field and exo-ionospheric density, interplanetary radio bursts, Jovian and terrestrial natural radio emission, and the galactic synchrotron spectrum. FARSIDE, and its precursor risk-reduction six antenna-node array PRIME, would be the first radio interferometers on the Moon. FARSIDE would break new ground by imaging radio emission from Coronal Mass Ejections (CME) beyond $2R_\odot$, monitor auroral radiation from the B-fields of Uranus and Neptune (not observed since Voyager), and detect radio emission from stellar CMEs and the magnetic fields of nearby potentially habitable exoplanets.

[1] Corresponding author jack.burns@colorado.edu.

[2] Center for Astrophysics & Space Astronomy, Department of Astrophysical and Planetary Sciences, University of Colorado Boulder, Boulder, CO 80516.

[3] NASA Goddard Space Flight Center, Code 695, Greenbelt, MD 20771.

[4] Department of Physics, University of California at Berkeley, Berkeley, CA 94720.

[5] Astronomy Department, MC 249-17, Pasadena, CA 91125.

[6] Department of Climate and Space Sciences and Engineering, University of Michigan, Ann Arbor, MI 48109.

1. **Introduction**

Even before Apollo 11's first human landing on the Moon, low radio frequency telescopes were being proposed for the lunar surface. At the first *Lunar International Laboratory Symposium* held in Athens in 1965, a Lunar Radio Astronomy Observatory was described which would eliminate restrictions on observations at ~kilometer wavelengths (~0.3 MHz) from the ground due to ionospheric and human-produced radio frequency interference (Gorgolewski 1966). This observatory was envisioned to have one antenna on the lunar surface and several more in lunar orbit to perform VLBI aperture synthesis. Proposed science goals included observations of the structure and spectra of extragalactic sources, 21-cm HI and OH line studies, planets, and HII regions, all observed during lunar night, and monitoring solar activity during the lunar day. This concept was prescient as many of these science goals remain viable today for low radio frequency science.

The first radio astronomy observations from the Moon were made by NASA's Radio Astronomy Explorer-2 (Explorer 49) satellite which collected data in lunar orbit from 1973-75. RAE-2 was placed around the Moon to shield the satellite from the Earth's surprisingly intense Auroral Kilometric Radiation (AKR), >40 dB brighter than the Galaxy below ~1 MHz (Alexander et al. 1975). RAE-2 carried a pair of long, 229- meter, travelling-wave V-antennas. RAE-2 focused on low radio frequency measurements of the Milky Way, solar and Jovian radio bursts, the AKR originating from the terrestrial magnetic field, and the extragalactic radio background.

Renewed interest in the Moon beginning in 1984 with the conference on *Lunar Bases and Space Activities of the 21$^{st}$ Century* brought forth new ideas for a lunar low frequency radio array (Douglas & Smith 1985) and a Moon-Earth baseline radio interferometer (Burns 1985). This was followed by additional low radio frequency mission concepts in workshops on *Future Astronomical Observatories on the Moon* (Burns & Mendell 1988), *Low Frequency Astrophysics from Space* (Kassim & Weiler 1990), *Science Associated with the Lunar Exploration Architecture* (NAC 2007), and more recently, *Discovering the Sky at the Longest Wavelengths with Small Satellite Constellations* (Chen et al. 2019). Other published lunar radio telescope mission concepts include Takahashi (2003), Jester & Falcke (2009), Lazio et al. (2011), Mimoun et al. (2011), Klein-Wolt et al. (2012), Zarka et al. (2012), and Bentum et al. (2020).

International landed payload missions have recently begun carrying low frequency radio instruments. The Chang'e-4 mission was the first to place a lander on the lunar farside. It housed a Very Low Frequency Radio Spectrometer and a tri-pole antenna, composed of three orthogonal, electrically-short, 5-m monopoles, operating between 0.1 and 40 MHz (Chen et al. 2019). Its science goals were to observe solar radio bursts during the lunar day and to measure the lunar ionosphere at its surface. Chang'e-4 also placed a communication satellite in a halo orbit about the Earth-Moon L2 Lagrange point and this satellite carried the Netherlands-China Low Frequency Explorer as a pathfinder for future lunar radio telescopes.

NASA's Payloads and Research Investigations on the Surface of the Moon (PRISM) program provides transportation for science instruments to the lunar surface as part of the agency's Commercial Lunar Payload Services (CLPS) initiative[7]. It is affording new access to the near and far sides of the Moon along with the poles. Two of the early CLPS landers will ferry low radio frequency telescopes to the lunar surface to investigate the lunar environs and the Sun, and to prepare for future radio telescopes on the lunar farside.

---

[7] https://www.nasa.gov/content/commercial-lunar-payload-services

This paper begins in Section 2 with an examination of the radio environment on the Moon based upon electrodynamic simulations and the best landing sites for radio telescopes. Then, the first NASA low frequency telescope experiments, ROLSES and LuSEE, are described in Section 3. In Section, 4 the results of a NASA-funded study of a Probe-class mission, FARSIDE, to robotically emplace an interferometric array on the Moon's farside later in this decade will be described; it is designed to investigate the Moon, planets. exoplanets, and the early Universe. A six-element precursor radio array, PRIME, is also described. Section 5 contains the Summary and Conclusions for this paper.

2. **The Lunar Radio Environment and Landing Sites**

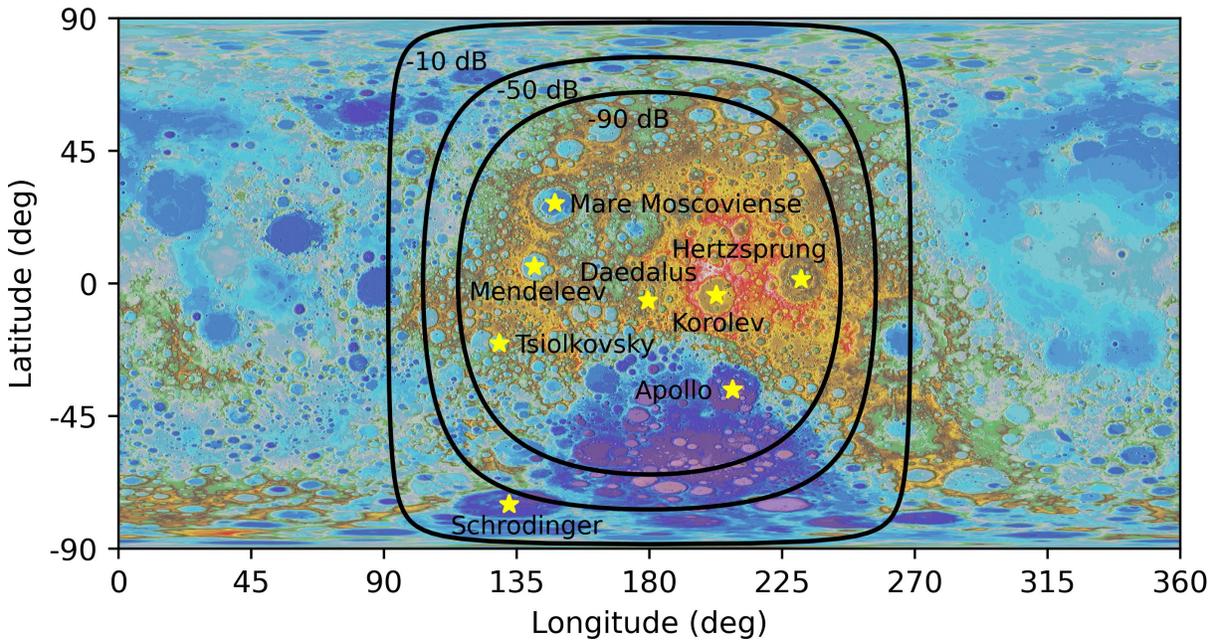

**Figure 1.** Map of RFI suppression at 100 kHz based upon numerical simulations from Bassett et al. (2020). Contours indicate suppression of -10, -50, and -90 dB relative to the incident intensity. Map colors indicate elevation. Potential landing sites are indicated by the yellow stars.

The lunar farside is unique in the inner solar system. It is free from human-made radio frequency interference (RFI) over much of its surface. The farside also protects astrophysical observations from the Earth's powerful AKR which occurs mostly below ≈0.5 MHz (e.g., Zhao et al. 2019). In addition, system noise at ≲1 MHz produced by electrons in the solar wind, which induce antenna currents (Meyer-Vernet & Perche 1989), are reduced by the lunar wake cavity especially at night (e.g., Farrell et al. 1998).

When choosing a landing site for a low radio frequency experiment, care should be taken that RFI and other contaminating radio emission is sufficiently attenuated to conduct the desired observations. While the farside is shielded from human-made interference over much of its surface, it is not perfectly radio-quiet due to the diffraction of low frequency waves around the limb of the Moon. In order to estimate the RFI intensity at various locations on the farside, Bassett et al. (2020) numerically simulated the propagation of low frequency radio waves around the Moon. By employing a finite difference time domain (FDTD) algorithm, commonly used for numerical electrodynamics problems, the amount of RFI suppression, taking into account the electromagnetic properties of the lunar material, can be calculated at any location on the farside. The code functions best at low frequencies (≲ 100 kHz) where the simulation resolution is comparable to the wavelength. Thus, the results of Bassett et al. (2020) can be used to evaluate the suitability of a landing site based upon the frequency band of a given instrument.

Due to the frequency-dependent nature of diffraction, the level of RFI suppression will be a function of frequency. Higher frequency waves will diffract less, leading to a lower level of RFI on the farside. In addition, the level of suppression at a given frequency will vary over the surface of the farside. As one moves from the lunar limb to the antipode of the Earth at the center of the farside, the intensity of RFI will decrease. This effect is shown clearly at a frequency of 100 kHz in Figure 1. As shown in the figure, the entirety of the farside is not completely radio quiet at 100 kHz. For example, portions of the South Pole Aitken Basin (SPAB; the large low-elevation, purple region in the lower center portion of the map) are likely to contain sufficient RFI to be detected by a sensitive instrument at 100 kHz. In contrast, at higher frequencies such as 10 MHz, the intensity of RFI is reduced by a factor of -90 dB (below which the results of numerical simulations are likely unreliable) within a few degrees of the limb such that the entire SPAB and nearly all of the farside are effectively free of RFI. Thus, the best and most radio-quiet regions on the Moon upon which to land a radio telescope that operates down to 100 kHz is inside the 90 dB contour. The FARSIDE array described in Section 4 will require a landing site flat to within roughly a 10-m elevation gradient over a 10-km diameter. There are numerous such locations with a few indicated in Figure 1.

### 3. Low Radio Frequency Experiments on NASA's Landed Payload Missions

The primary goal of both NASA landed payload radio science experiments, ROLSES and LuSEE, is a proof-of-concept that good, clean radio measurements can be made on the Moon. Adherence to a strong electro-magnetic compatibility plan will determine whether high fidelity spectra are possible from the lunar surface. This includes both the instruments and the landers.

### 3.1 ROLSES – Radio wave Observations at the Lunar Surface of the photoelectron Sheath

The low-frequency ROLSES radio receiver system for the lunar surface will provide radio spectra from 10 kHz – 30 MHz with ~4 second resolution. ROLSES will have two frequency bands, from 10 kHz – 1 MHz and 300 kHz – 30 MHz. The two frequency bands will each be processed using a 512-bin real-time digital Weighted-Overlap and Add (WOLA) spectrometer with 512 bins. The low band resolution will be 1.76 kHz, and the high band resolution will be 58.01 kHz. Because we are limited to a data rate on NOVA-C of 17 kbps, the time resolution will be averaged spectra from two antennas for 4 seconds, and then averaged spectra from the other two antennas for the next 4 seconds, so that the effective time resolution is essentially 4 seconds. These Radio wave Observations at the Lunar Surface of the photoElectron Sheath (ROLSES) will permit determination of the electron photo-sheath density from ~1 to ~3 m above the lunar surface. This is of scientific value and is also important to determine the effect on the antenna response of larger lunar radio observatories with antennas laying on the lunar surface. The ROLSES data also relate to other science goals as described below. In addition, the data will provide continuous spectra of radio frequency interference from terrestrial transmitters for the 14-day mission, which is important information to confirm how well a near side lunar surface-based radio observatory could observe and image solar radio bursts in the frequency range of 0.01 – 30 MHz for the first time. This frequency range corresponds to distances from the Sun of ~1 AU to ~1 solar radii, which is a very interesting spatial volume for future imaging of solar radio bursts, to better understand the emission mechanisms and to enhance their space weather prediction applications.

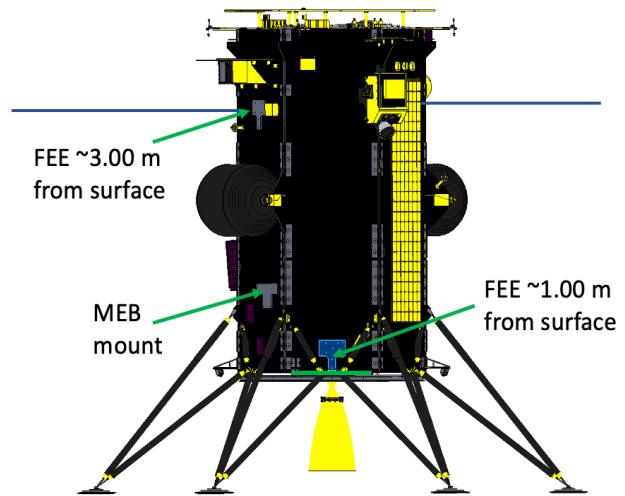

**Figure 2.** Diagram of the NOVA-C lander that will take the ROLSES payload to the lunar surface in 2021. Locations of the Main Electronics Box and 2 of the 4 Front End Electronics units are indicated; horizontal lines represent 2 STACER antennas.

ROLSES is one of the payloads to be taken to the lunar surface by the NOVA-C lander of Intuitive Machines LLC[8], which is funded by NASA. The ROLSES payload consists of a major electronics box (MEB) and four STACER (Spiral Tube & Actuator for Controlled Extension/Retraction; Bale et al. 2008) antennas and front-end electronics units. Two of the STACERs are mounted on NOVA-C on opposite sides of the lander at ~1 m above the lunar surface, and the other two are mounted on NOVA-C close to ~3 m above the lunar surface, on opposite sides of the lander and oriented at azimuths perpendicular to the ~1 m height STACER units. Figure 2 shows most of these components.

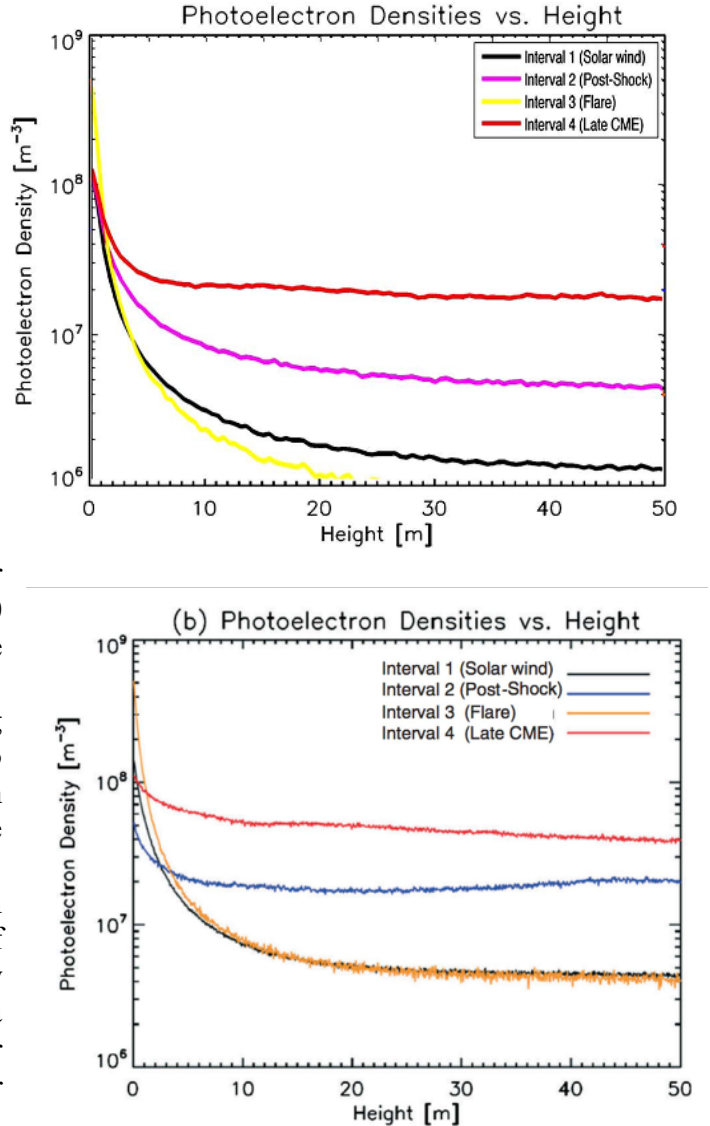

**Figure 3**. The photoelectron density as a function of height above the lunar surface is indicated in the two plots from the simulation codes by Zimmerman et al. (2011) and by Poppe & Horanyi (2010) for various solar wind environments. At 1-m height for a typical solar wind, the plasma density is ~$5\times10^7$ m$^{-3}$ and the electron plasma frequency ($f_{pe}$) is ~ 64 kHz (Farrell et al. 2013).

---

[8] https://www.intuitivemachines.com/lunarlander

### 3.1.1 ROLSES Science goals

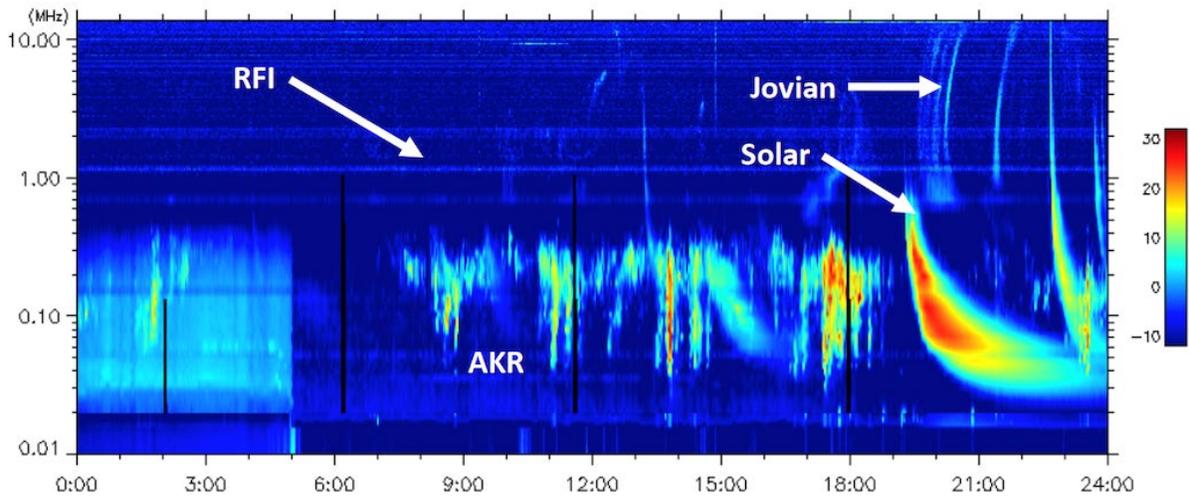

**Figure 4**: The WAVES instrument on the Wind spacecraft (Bougeret et al. 1995) at Sun-Earth L1 shows solar radio bursts, Earth's auroral kilometric-wavelength radio bursts (AKR), terrestrial ground-based transmitters (RFI), and Jovian radio emissions, during a 24 hr interval on 2/20/2012. ROLSES will be capable of making similar observations, which would demonstrate some of the science data provided by the lunar radio array, FARSIDE, described in Section 4.

On the lunar dayside, photoelectrons are quasi-constantly emitted from the Moon's surface and this electron flux acts to typically charge the dayside lunar surface a few volts positive. In arriving at an equilibrium surface potential, the surface will charge to balance the two primary currents: the outgoing photo-electron flux, $J_p$, against the incoming solar wind electron thermal flux, $J_e$. In nominal solar wind conditions, $J_p > J_e$ and the surface charges positive, trapping most of the photoelectrons (Farrell et al., 2013). The photoelectrons are related to surface charging, as well as the charging of exploration vehicles on the lunar surface. They also play a role in the mobility of lunar dust particles, which consequently varies based on the solar wind conditions. Therefore, the best possible understanding of the photoelectron sheath is important for future exploration of the lunar surface. There are several simulation code results published for the lunar photoelectron sheath. Figure 3 shows two plots from such simulations showing the significant increase in the electron density inside a height of 10 m. Note also that these two plots, which are simulations for the same interval (May 1-3, 1998), do not have exactly the same results, which ROLSES is designed to resolve.

The sheath will impact radio observations on the order of 100 kHz and attenuate radio waves at lower frequencies. Of course, variations in the solar wind have a significant effect on the photoelectron sheath, as shown in Figure 3. Assuming that these models are correct, then the good news is that an astrophysics-oriented radio observatory could make daytime observations down to less than 1 MHz without any impact from the electron sheath. On the other hand, any observatory making lower frequency observations, such as a heliophysics solar radio burst imaging observatory would likely have refraction and attenuation issues for observations at less than 100 kHz due to the photoelectron sheath. To better understand this situation, ROLSES will make measurements of the photoelectron sheath density scale height and determine the frequency at which the incoming radio waves will be attenuated/cutoff. ROLSES measures these values based on a thermal noise spectrum, possible wave activity at the electron plasma frequency (fpe), and absorption of the radio spectra of remote radio sources at frequencies at and below fpe. Figure 4 shows examples of the incoming radio waves that ROLSES will be detecting.

Spacecraft in the interplanetary environment or orbiting planets may be struck by dust particles, which releases electrons and ions from the surface, affects the surface photoelectron environment, and creates signals that are detectable. ROLSES might detect dust impacting NOVA-C lander in a similar way. The time resolution of 4 sec does not permit detecting individual dust particles, but could detect dust "clouds", like Figure 5.

There is no ground plane below the ROLSES antennas, so some detected radio waves will likely penetrate the lunar surface. They may be reflected at some depth, and ROLSES may be able to detect such reflection, assuming its duration is at least 4 sec, providing some information about the structure below the lander. Previously, the Apollo 17 lunar Surface Electrical Properties (SEP) instrument made such measurements at 6 frequencies, with a signal generator that sent radio waves down to a few km into the Moon. Grimm (2018) describes recent analysis of the SEP results in detail, e.g., "Because no deep interfaces were detected, the thickness of the Taurus-Littrow volcanic fill must exceed 1.6 km and possibly 3 km."

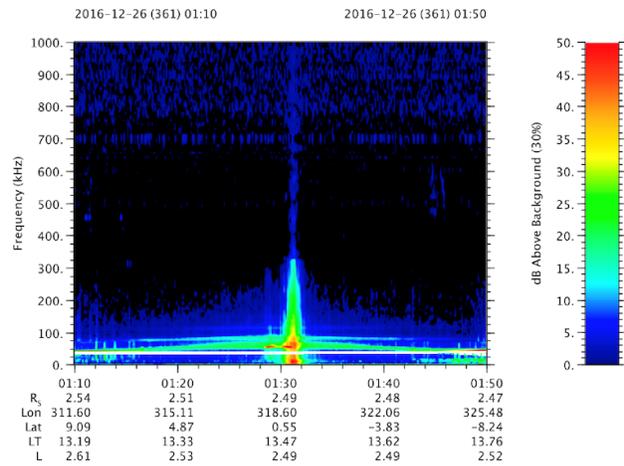

**Figure 5**: The plot shows the dust signal detected by the Cassini spacecraft when crossing the Saturn F-ring.

### 3.1.2 Technology goal

ROLSES will provide continuous spectra of radio frequency interference (RFI) from terrestrial transmitters (Figure 6) for the ~14-day mission; information to confirm how well a near-side lunar surface-based radio observatory could observe and image solar radio bursts in the frequency range of 0.01 to ~10 MHz for the first time. Its deployed STACER lengths are 2.5 m. Its frequency range of 10 kHz – 30 MHz, is divided into two bands: 10 kHz – 1 MHz and 300 kHz – 30 MHz. The low frequency band resolution is ~1.76 kHz, and the high frequency band resolution is ~58.01 kHz. The low-band sensitivity is expected to be ~3 $\mu$V/Hz, which will be further refined after the lander and system noise are fully defined.

Like all lunar surface payloads, the ROLSES electronics need to survive the maximum lunar surface temperatures. To deal with this, each of the ROLSES components has a radiator plate attached to it to keep temperatures below ~65 C. This will permit it to function for the complete 14-day NOVA-C mission. NOVA-C receives the data from ROLSES and sends it to the ground stations. Currently the NOVA-C mission is targeting landing on the Moon in Q4 of 2021. Intuitive Machines has chosen the landing site to be in Oceanus Procellarum near Vallis Schröteri (Schroter's Valley), in part because there should be many smooth surface areas there. The location relative to the Earth-Moon line is 25° N, 50° W. The location is reasonable for the ROLSES payload, but not unique.

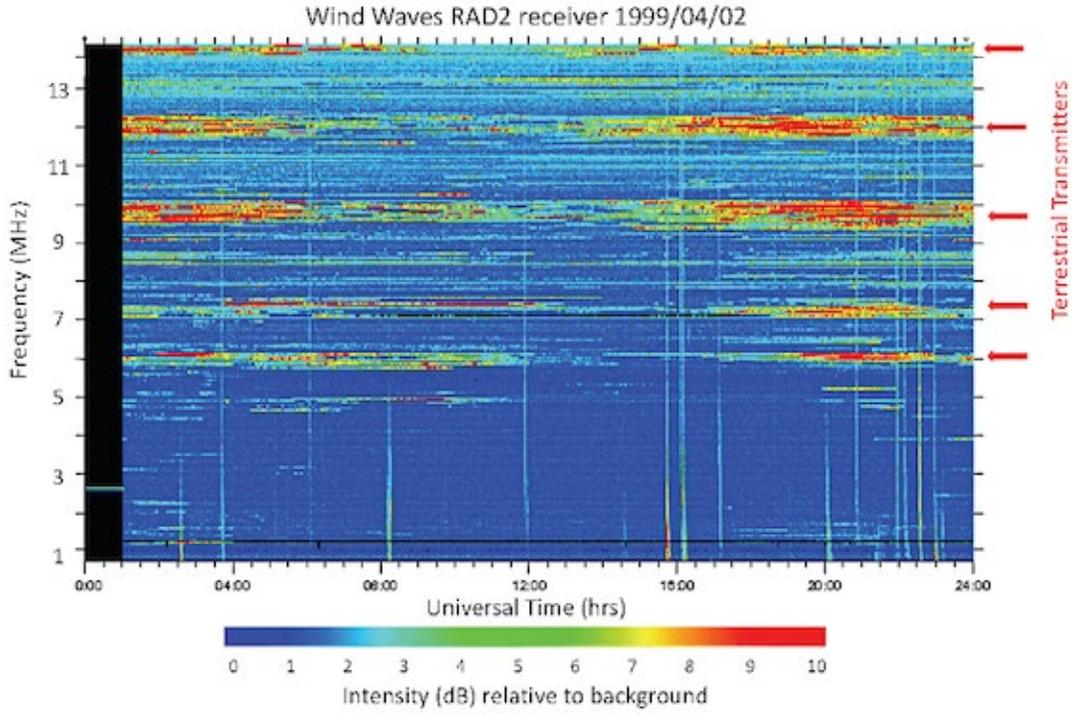

**Figure 6:** Dynamic spectrum shows terrestrial RFI observed by Wind/WAVES on 1999/04/02 when it passed the Moon.

### 3.2 LuSEE - The Lunar Surface Electromagnetic Experiment

The Lunar Surface Electromagnetics Experiment (hereafter LuSEE) is a suite of sensors and electronics designed to make comprehensive measurements of the electromagnetic environment of the lunar surface from DC to radio frequencies (~20 MHz). LuSEE was selected by NASA in 2019 through the CLPS initiative. As of this writing, NASA intends to place LuSEE on the lunar farside within the Schrödinger Basin in 2024.

#### 3.2.1 LuSEE Landing Site

While the Schrödinger Basin (see Fig. 1) is located on the lunar farside, it will not be completely radio quiet near the low end of the LuSEE frequency band. In fact, at 100 kHz, the level of suppression of radio interference will vary significantly across the basin. As shown in Figure 7, the level of attenuation varies from a factor of approximately 25 dB at the southwest corner of the basin to near 55 dB at the northeast corner. Based on these levels of suppression, it is likely that RFI and AKR will be present at a sufficient level to be detected by LuSEE. Specifically, the AKR from Earth's dayside will be observed to be comparable to the flux density of the Galaxy at 100 kHz (~105 Jy) whereas the AKR from Earth's nightside will be ~40 dB brighter than the Galaxy in the SW quadrant of Schrödinger (see Fig. 2 of Zarka et al. 2012).

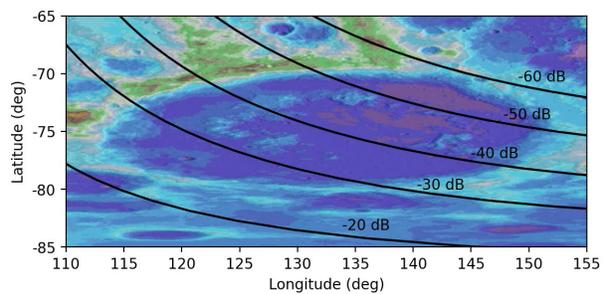

**Figure 7.** Map of RFI suppression at 100 kHz in the Schrödinger Basin based on Bassett et al. 2020. The labels in the figure indicate the level of suppression for each contour.

However, the presence of radio interference in the Schrödinger Basin provides a method with which to test predictions of the radio environment. In addition to its primary science goals, LuSEE will provide a ground-truth measurement with which to compare the results of the numerical simulations of Bassett et al. (2020). Further work is underway to extend the results of Bassett et al. (2020) to include a tenuous lunar ionosphere and electron sheath, which will likely influence the propagation of waves at frequencies as low as 100 kHz. These simulations, in conjunction with LuSEE measurements, will provide the most comprehensive study of the radio environment of the farside, and specifically the Schrödinger Basin, to date.

In contrast to the radio interference at lower frequencies, LuSEE high band measurements at frequencies above 10 MHz will be shielded from terrestrial (artificial) shortwave emissions, which are attenuated by a factor of at least 90 dB over the entire basin. This lack of contamination makes the Schrödinger Basin an excellent landing site for LuSEE to make precision observations of the radio sky, as detailed in the following section.

### 3.2.2. LuSEE Science Goals

The science targets of the LuSEE suite include lunar surface plasma physics, the science of dust transport near the surface, and to serve as a low frequency radio pathfinder mission. The lunar surface plasma environment is relatively poorly understood, with a true paucity of measurements. However, any surface-landed asset (human or robotic) must function in the surface plasma environment. The surface electric potential that will control the transport of (charged) dust should be highly variable, depending on the upstream plasma parameters, solar UV irradiance, solar zenith angle, and local shadowing (Reasoner and Burke, 1972) with predictions in the range of a few Volts to ~100 V (Poppe et al. 2012; Stubbs et al., 2014). LuSEE will make true measurements of the DC electric potential using a current-biased double-probe technique, which is appropriate for electric field measurements in low density plasmas (e.g., Mozer et al. 1983). The Moon is thought to sustain a tenuous "exo-ionosphere" supported by photoionization, electron impact ionization, and charge exchange. Radio occultation measurements have suggested rather large plasma densities 500-1000 cm$^{-3}$ with scale heights of tens of km (e.g. Vasiliev et al. 1974), while in-situ measurements suggest smaller surface plasma density. The THEMIS-ARTEMIS mission measured electron plasma waves which suggested an ionospheric density of ~0.1-0.3 cm$^{-3}$ (Halekas et al. 2018). LuSEE is designed to make sensitive measurements of electron plasma waves/noise using a technique called "quasi-thermal noise spectroscopy" (e.g. Meyer-Vernet and Perche, 1989) and can measure local electron density and its variability, and in some cases the electron temperature, on the lunar surface. At higher frequencies, up to ~20 MHz, LuSEE will make sensitive radio frequency measurements of interplanetary radio bursts, Jovian and terrestrial natural radio emission, and the galactic synchrotron spectrum. LuSEE will provide best-yet spectral measurements of the quiet radio sky from the lander surface.

The LuSEE experiment is derived from flight spare and engineering model hardware from the NASA Parker Solar Probe "FIELDS" instrument (Bale et al., 2016) (hereafter PSP/FIELDS), the NASA STEREO WAVES (Bougeret et al., 2008; Bale et al, 2008) (S/WAVES), the NASA Van

Allen Probes Electric Field and Waves (EFW) experiment (Wygant et al 2013), and the ESA Solar Orbiter Radio and Plasma Waves (RPW) suite (Maksimovic et al, 2020).

The LuSEE system consists of a 'Main Electronics Package' which is comprised of three digital signal processing engines, a low voltage power converter, and data processing units with interfaces to the spacecraft command and data-handling system. These units are all flight spares, or nearly build-to-print, from the PSP/FIELDS instrument on-orbit and have demonstrated excellent scientific capability. The Radio Frequency Spectrometer (RFS) subsystem is a dual channel, baseband digital radio receiver that operates in two frequency regimes (10 kHz – 2 MHz and 1.3 – 19.2 MHz), and is described by Pulupa et al. (2017). The RFS uses polyphase filter techniques to reject out-of-band noise sources and is highly configurable. Early results from the RFS on PSP (Pulupa et al. 2020) show superb sensitivity and capability, resolving the galactic spectrum down to a few hundred kHz (and a few nV). The PSP/FIELDS RFS is arguably the most capable LF radio instrument ever put into orbit and LuSEE inherits that capability. The LuSEE Digital Fields Board (DFB) measures DC- and AC-coupled voltage inputs from the antennas and search coil magnetometer (Malaspina et al., 2016) and produces time-series measurements to 150 k-Samples/s and spectral and cross-spectral matrices to 75 kHz. DFB is highly configurable and uses a triggered 'burst mode' to capture impulsive signals. The LuSEE preamp-DFB system has capabilities to measure DC voltage perturbations to +/- 100 V, producing direct measurements of the lunar surface potential. The LuSEE Time Domain Sampler (TDS) is a snapshot waveform sampler that measures 5 analog channels (and one digital) at speeds up to 2 M-Samples/s and some accumulated statistics associated with these waveforms (Bale et al., 2016). The TDS on PSP/FIELDS has produced spectacular results on the distribution of interplanetary dust (Page et al. 2020; Malaspina et al. 2020; Szalay et al. 2020) and the physics of spacecraft/dust interactions (Bale et al. 2020). A central processing unit, the Data Controller Board and low-noise power converter round out the complement in the MEP.

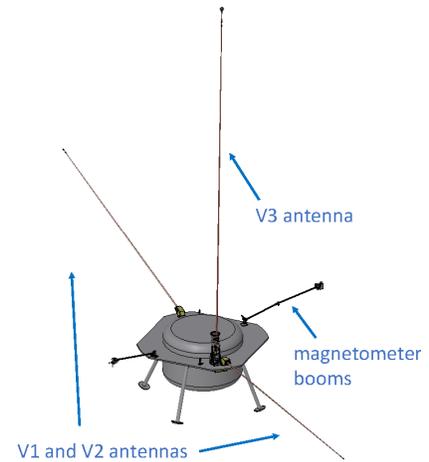

**Figure 8.** This schematic shows the LuSEE deployable sensors: V1 and V2 are 6-m STACER antennas deployed to form a dipole. V3 is a stacer with a spherical voltage probe used to measure the vertical electric field. Fluxgate and search coil magnetometers are mounted on 2-m and 1-m carbon fiber booms, respectively.

LuSEE will use 3 voltage probes and 2 boom-mounted magnetometers as sensors. Figure 8 shows the LuSEE sensors mounted on a strawman concept lander. The V1 and V2 antennas are flight spare units from the STEREO/WAVES instrument (Bougeret et al., 2008). The V1/V2 antennas are 6-m long, ~1" diameter BeCu STACER antennas (Bale et al., 2008); STACERs are cold-rolled metal springs that deploy under their own spring force (no motor) and have been used on hundreds of space projects, with no known failures. The V1/V2 antennas will be deployed out away from the spacecraft and at an angle to the horizontal so that they sag in lunar gravity to form a dipole, which has been modeled mechanically. The V1/V2 dipole will be the primary measurement for the RFS/radio frequencies. The V3 antenna is a flight spare unit from the EFW experiment on the Van Allen Probes mission (Wygant et al. 2017) and also uses STACER technology. However, V3 holds a small spherical voltage probe that is current-biased and will be used to measure the DC voltage between V3 (at 6+m above the surface) and the V1/V2 pair, yielding a measurement of the vertical electric field. Magnetic field measurements on LuSEE will be made from a 3-axis fluxgate magnetometer (MAG), derived from PSP and MAVEN heritage and a 3-axis search coil magnetometer (SCM) flight spare from the Radio and Plasma Waves instrument (Maksimovic et al. 2020) on the ESA Solar Orbiter mission. The MAG sensor measures magnetic fields with < 1 nT resolution at cadences to 293 Samples/s, while the SCM has no DC sensitivity, but otherwise excellent response to 10s of kHz.

### 4. The FARSIDE Low Radio Frequency Array

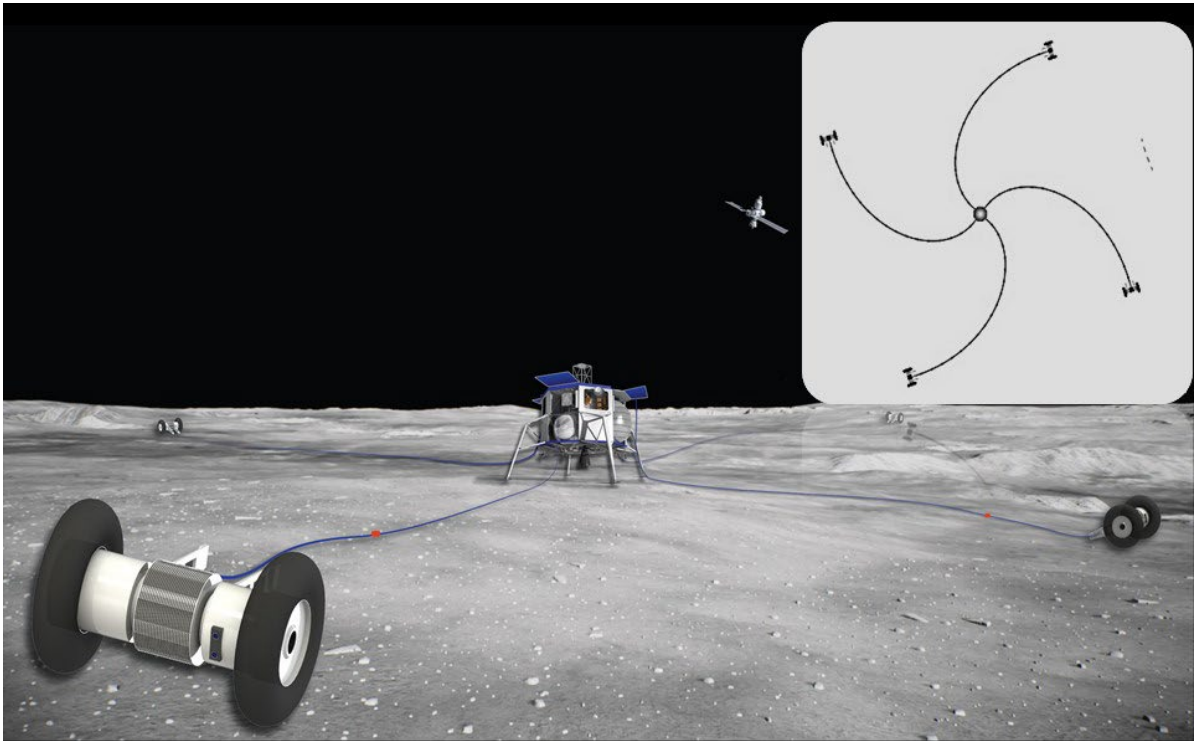

**Figure 9:** FARSIDE consists of three components: a commercial lander carrying the base station, four single-axel rovers to deploy antenna nodes, and a 128x2 (two orthogonal polarizations) node antenna array. The array will be deployed in a spiral pattern as shown in the upper right. Tethers connect the base station to the nodes, providing communications and power. NASA's lunar-orbiting Gateway may serve as a data relay to Earth.

ROLSES and LuSEE are paving the road ahead, demonstrating the viability of high dynamic range spectral and polarimetric low radio frequency observations on the Moon. As a next step, NASA recently funded the first engineering design study of an interferometric array for possible deployment by a large CLPS lander on the lunar farside later in this decade (Burns et al. 2019a). FARSIDE[9] is a potential example of NASA Astrophysics' proposed new mission class called Probes which are PI-led missions with a funding cap of ≈$1 billion, similar to that for the Planetary

---

[9] FARSIDE is an acronym for **F**arside **A**rray for **R**adio **S**cience **I**nvestigations of the **D**ark ages and **E**xoplanets. The Dark Ages science is not presented here but is described in Burns et al. (2019a).

Science's New Frontier class. The FARSIDE architecture and operations concept were developed via a series of studies at JPL. These studies assessed the technical feasibility of the instrument and the mission, and the cost-realism for a Probe-class mission. The current FARSIDE design architecture consists of 128 non-cospatial orthogonal pairs of antenna/receiver nodes distributed over a 10-km diameter in a four-arm spiral configuration shown in Figure 9. The nodes are deployed in zig-zag pattern with the first antenna of the pair having a vertical polarization axis (Ey) and the next having a horizonal axis (Ex), producing measurements of all four Stokes parameters. The FARSIDE payload mass is estimated to be 1750 kg which would be delivered to the lunar surface by a commercial lander, such as Blue Origin's Blue Moon lander[10]. Four single-axel teleoperated rovers would deploy the nodes which are connected to the base station (lander) by science tethers, providing communications, data relay, and power.

The array is planned for operations over five years with a frequency range of 0.1-40 MHz. At its lowest frequency, FARSIDE would operate two orders of magnitude below the Earth's ionospheric cutoff. The array would image the sky visible from the lunar surface once per minute to search for radio bursts from energetic particle events within the solar system (e.g., Type II and III solar bursts; CME-generated planetary auroral emissions) and in exoplanetary systems.

A summary of the instrument and its sensitivity along with the overall mission design is presented next in Section 4.1. The focus then turns to the solar and extrasolar system science with this array in Section 4.2, including measurements of Type II/III radio bursts from the Sun, auroral radio radiation from the outer planets including the ice giants Uranus and Neptune, space weather and magnetospheres in exoplanetary systems, and probing the lunar mega-regolith layers. In Section 4.3, a six-element prototype radio interferometer, PRIME, is introduced which would reduce risks for the larger FARSIDE array.

### 4.1 Overview of the FARSIDE Instrument and Mission Concept

Details on the engineering design for FARSIDE are found in the NASA mission concept study report (Burns et al. 2019a). Here, a high-level overview of the instrument is given with a few updates that extend the concept study.

#### 4.1.1 FARSIDE Sensitivity

Beginning with the distinctive radio-quiet characteristics of the farside described in Section 2 as a requirement, an astronomical array at frequencies 0.1-40 MHz is driven by sensitivity and spatial resolution. The sensitivity is determined by the number and type of antennas used in the array and the resolution is governed by the maximum baseline. For FARSIDE, we use 100-m tip-to-tip thin-wire antennas. It was determined that 128x2 dipole antennas will provide the needed sensitivity to span the science cases described in Section 4.2, especially probing the space plasma weather and magnetospheres in exoplanetary systems. With an array diameter of 10-km, FARSIDE achieves a resolution of 10° FWHM at 200 kHz and 10 arcmin at 15 MHz.

Sensitivity specifications for FARSIDE are summarized in Table 1. The array sensitivity, or minimum detectable flux density, in Table 1 depends upon the frequency bandwidth, the integration time, the system temperature, and the effective area of the array. The effective area is determined by the dipole length and the antenna impedance. This was modeled using NEC4.2 simulations which take into account that the dipole wires rest directly on the regolith. The system temperature depends upon the sky and regolith temperatures, and the front-end amplifier. See Burns et al. (2019a) for complete details of the modeling and simulations.

---

[10] https://www.blueorigin.com/blue-moon/

**Table 1.** FARSIDE Sensitivity

| Quantity | Value |
|---|---|
| Frequency Coverage | 0.1 – 40 MHz (1400 × 28.5 kHz channels) |
| Antenna Efficiency | $6.8\times10^{-6}$ @ 200 kHz; $9.5\times10^{-5}$ @ 15 MHz |
| System Temperature ($T_{sys}$) | $1\times10^6$ K @ 200 kHz; $2.7\times10^4$ K @ 15 MHz |
| Effective Collecting Area ($A_{eff}$) | ~12.6 km$^2$ @ 200 kHz; 2240 m$^2$ @ 15 MHz |
| System Equivalent Flux Density ($2k_BT_{sys}/A_{eff}$) | 230 Jy @ 200 kHz; $2.8\times10^4$ Jy @ 15 MHz |
| 1$\sigma$ Sensitivity for 60 seconds, $\Delta\nu=\nu/2$ | 66 mJy @ 200 kHz; 933 mJy @ 15 MHz |

The FARSIDE instrument front end uses the electrically short antennas to achieve sky background noise-limited observations. On the lunar farside, the lunar highlands are thick, have low conductivity, and vary slowly with depth, thus removing the need for a ground plane. The net impact of this multipath effect is to introduce a direction dependent component to the array synthesized beam. FARSIDE will use an orbital beacon for calibration and to map both the antenna beam pattern and the array synthesized beam to fully capture this effect. The 100-m antennas are embedded within the tether and will be placed directly on the lunar surface. The array leverages existing designs from high heritage front-end amplifiers, fiber optics, and the correlator system based upon ground-based observatories such as the Owens Valley Radio Observatory Long Wavelength Array (OVRO-LWA, Anderson et al. 2018).

**4.1.2 Deployment and Operations**

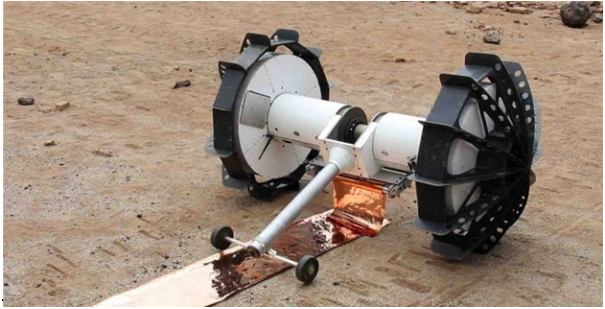

**Figure 10:** This JPL Axel rover is shown deploying a polyimide tether similar to that proposed for FARSIDE. See Matthews & Nesnas (2012) for rover details.

The concept design calls for operations during the lunar night as well as the lunar day using proven low temperature electronics and batteries. Recent tests indicate that core constituents such as resistors and capacitors, forming the basis of the electronic components of the FARSIDE receiver nodes, are capable and survivable down to temperatures <-180 C (Ashtijou et al. 2020).

The 128 antenna node pairs are distributed along four spirals, as shown in the Figure 9 insert. This configuration with four JPL Axel rovers (Figure 10) would allow the full deployment of the array in one lunar day. A tether connects the nodes to each other and to the central base station for data transmission and power. For each spiral arm, the rover would carry a set of antenna nodes from the base station, unspooling the tether, before returning to the base along a different path. We assume that the landing site and rover path are relatively flat, free of obstacles with scale sizes ≳1-meter.

The base station includes solar panels and batteries to provide power for the receivers and the signal processing using an FX correlator (e.g., Kocz et al. 2014), as well as communication of the data back to Earth, possibly via NASA's lunar Gateway. The rovers draw power along the tether from the base station. The rover will either be operated from Earth (with ≈5 second latency) or teleoperated via astronauts aboard the lunar Gateway (with ≈0.5 second latency) (Burns et al. 2019b). After cross-correlation via the FX correlator, visibility data will be conveyed to Earth for further analysis. This is estimated to be about 130 GB of data per 24 hour period. The goal is to process 10,000 square degrees of the sky every 60 seconds over 1400 frequency channels. This would then produce dynamic spectra at the location of each bursting object including Type II/III solar emissions and bursts in any of the several thousand nearby exoplanetary system counterparts as well as CME-driven auroral emissions from nearby terrestrial exoplanets.

### 4.1.3 Synthetic FARSIDE Observations

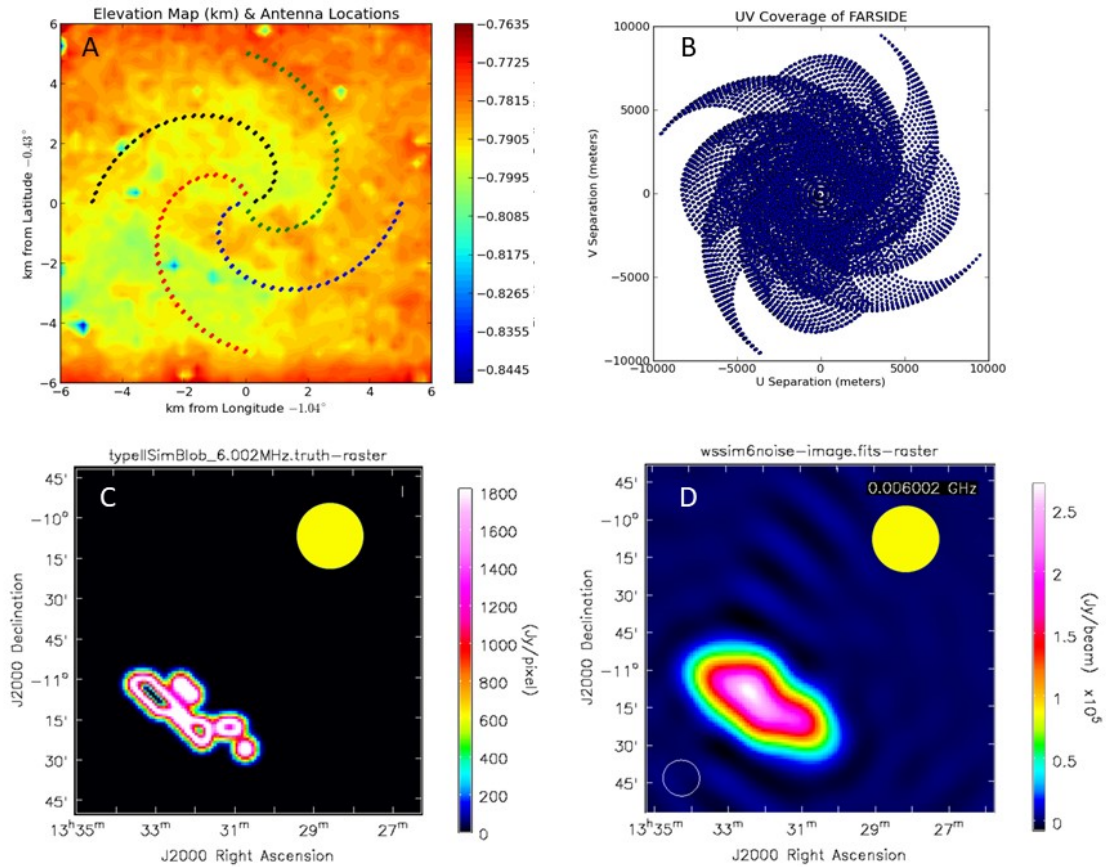

**Figure 11.** A: Distribution of FARSIDE antennas on the lunar surface. Elevations derived from NASA LRO LOLA. B: Instantaneous u-v coverage for the antenna distribution shown in A. C: Estimated emission from a solar CME shock front (Type II radio burst) at 6 MHz. Yellow disk is approximate size and location of the Sun. D: "Cleaned" radio map of Type II burst in C as imaged via the FARSIDE array at 6 MHz with 1 minute integration. White circle in the lower left is the FWHM of the synthesized beam.

Recently, Hegedus et al. (2020) utilized a combination of the CASA software package (McMullin et al. 2017), lunar surface maps from LRO LOLA (Barker et al 2016), and Spice (Acton 1996) to define a simulated array on the Moon's surface. This approach properly aligns the reference frames from the lunar surface to the sky for the purpose of observing low frequency emissions. Here, the software pipeline has been adapted for the simulation of FARSIDE observations. The 10 km, four-armed spiral pattern of FARSIDE on top of an elevation map from LOLA data is shown in panel A of Figure 11. The array consists of 256 antennas, 128 pairs of X and Y polarizations that lie nearly side by side, each of which are 100-m in length. The instantaneous u-v coverage (projected baselines) of the array for an observation at its zenith is shown panel B of Figure 11.

One of the scientific targets of FARSIDE is imaging solar Type II and III radio bursts. The simulation pipeline for FARSIDE has been applied to this problem, using a model-inspired Type II burst as input to the array in order to assess its reconstructive capabilities. This model Type II burst is shown in panel C of Figure 11. This model emission was obtained by taking a subset of data from an MHD recreation of the radio-loud, Earth-directed coronal mass ejection (CME) observed on 5/13/2005 (Manchester et al 2014). A subset of the data was identified via applying parameter-based thresholds on the simulation data for each timestep in the simulation. The first threshold used for these data was that the regions included must have an increase in local entropy of at least a factor of 4 compared to the pre-shocked data. This entropy threshold generally captured the location of the shock within the simulation. The other threshold included regions with a local de Hoffmann-Teller speed of at least 2250 km/s. This is the speed in the frame where the upstream plasma bulk velocity and magnetic field are parallel, and therefore there is no upstream convective (V×B) electric field. This de Hoffmann-Teller speed was identified by Pulupa et al (2010) to be most highly correlated with in-situ observations of Langmuir waves from accelerated electrons at 1 AU coinciding with shocks, agreeing well with the predictions of fast Fermi acceleration theory.

Panel C shows the projected sky distribution after applying the above parameter-based thresholds, illustrating only the regions that have upstream plasma frequencies 6-6.1 MHz. This map is then slightly blurred by applying a Gaussian kernel with a full width at half maximum (FWHM) of 0.07°. The resulting "truth image" is then normalized to a brightness of $10^6$ Jy and fed into the simulated FARSIDE array. The simulated response of the FARSIDE antenna on the lunar surface assumes $A_{eff}/T_{sys}$ = 0.065 at 6 MHz, consistent with the array parameters in Table 1. Appropriate thermal noise was added to create a more realistic response of FARSIDE to the model Type II emission. These data were then imaged and "cleaned" (i.e., deconvolved) using the widefield imaging software WSClean (Offringa et al 2014), with the resulting image shown in panel D of Figure 11. This simulation pipeline confirms that FARSIDE would provide unprecedented imaging capabilities for low frequency solar radio bursts, and could help explain how energetic particles are being accelerated at CME driven shocks.

## 4.2 Solar System and Exoplanetary Radio Science with FARSIDE

### 4.2.1 Heliophysics

During the solar illumination of lunar day, FARSIDE will observe solar radio bursts, which are one of the primary remote signatures of electron acceleration in the inner heliosphere. FARSIDE extends the spectroscopic observations of LuSEE and ROLSES by producing the first radio images of Type II and Type III radio bursts at $\gtrsim 2R_\odot$ (Figure 11). Type II bursts originate from suprathermal electrons (E > 100 eV) produced at shocks. These shocks are generally produced by CMEs as they expand into the heliosphere with Mach numbers >1. Emission from a Type II burst drops slowly in frequency as the shock moves away from the Sun into lower density regions at speeds of 400–2000 km s$^{-1}$. Type III bursts are generated by fast (2–20 keV) electrons from magnetic reconnection, typically due to solar flares. As the fast electrons escape at a significant fraction of the speed of light into the heliosphere along open magnetic field lines, they produce emission that drops rapidly in frequency. As discussed in §4.2.3, FARSIDE will also attempt to detect intense type II and type III emissions from other stars in the solar neighborhood.

*Acceleration at Shocks:* Observations of CMEs near Earth suggest electron acceleration generally occurs where the shock normal is perpendicular to the magnetic field (Bale et al. 1999), similar to acceleration at planetary bow shocks and other astrophysical sites. This geometry may be unusual in the corona, where the magnetic field is largely radial. There, the shock at the front of a CME generally has a quasi-parallel geometry. Acceleration along the flanks of the CME, where the magnetic field-shock normal is quasi-perpendicular would seem to be a more likely location for the electron acceleration and Type II emission. FARSIDE has sufficient resolution (Figure 11) to localize these acceleration site(s), yielding the preferred geometry of the shock normal relative to the magnetic field direction for radio emission around CMEs.

*Electron and Ion Acceleration:* Observations at 1–14 MHz made with the Wind spacecraft showed that complex Type III-L bursts are highly correlated with CMEs and intense (proton) solar energetic particle (SEP) events observed at 1 AU (Cane et al. 2002; Lara et al. 2003; MacDowall et al. 2003). While the association between Type III-L bursts, proton SEP events, and CMEs is now secure, the electron acceleration mechanism remains poorly understood. Two competing sites for the acceleration have been suggested: at shocks in front of the CME or in reconnection regions behind the CME. For typical limb CMEs, the angular separation of the leading edge of the shock and the hypothesized reconnection region behind the CME is approximately 1.5° when the CME shock is 3–4 $R_\odot$ from the Sun.

*CME Interactions and Solar Energetic Particle (SEP) Intensity:* Unusually intense radio emission can occur when successive CMEs leave the Sun within 24 hours, as if CME interaction produces enhanced particle acceleration (Gopalswamy et al. 2001, 2002). Statistically associated with intense SEP events (Gopalswamy et al. 2004), this enhanced emission could result from more efficient acceleration due to changes in field topology, enhanced turbulence, or direct interaction of the CMEs. Lack of radio imaging makes it difficult to determine the nature of the interaction. Images with ~2° resolution would give Type II locations and permit identification of the causal mechanism and the relation to intense SEPs.

FARSIDE provides the requisite spatial resolution, as well as exquisite signal-to-noise and imaging capability (due to large number of unique baselines) spanning the 1–14 MHz band observed by the Wind spacecraft and extending much wider at both lower and higher frequencies. It will be a uniquely powerful tool for understanding the complex interplay of CMEs, SEPs and associated shocks and particle acceleration.

### 4.2.2 Outer Solar System Planets

In 1955, the planet Jupiter was serendipitously detected to be a source of bright, highly variable radio emission at decametric wavelengths (Burke and Franklin 1955). This radio emission was manifested as intense, highly polarized bursts, so bright as to often outshine the Sun at these low frequencies (<40 MHz). Bursts were only detectable over specific ranges of rotational phase of the planet, with certain flavors also found to be strongly modulated by the orbit of the volcanic moon Io (Bigg 1964). This discovery revolutionized our understanding of planetary magnetospheric physics, providing the first direct confirmation of the presence, strength, and extent of the Jovian magnetosphere. In particular, the radio bursts, generally attributed to electron cyclotron maser emission, were found to be produced at the electron cyclotron frequency, $\nu_{ce} \approx 2.8\,B$ MHz, where $B$ is the magnetic field strength (in Gauss) at the source of the burst, and thus enabled direct determination of magnetic field strength.

All the magnetized planets in our solar system, including Earth, have since been found to produce similar bright coherent radio emission at low frequencies, predominantly originating in high magnetic latitudes and powered by auroral processes. It is notable that only Jupiter, with a maximum polar magnetic field strength of 14 G, produces radio emission that can penetrate through the Earth's ionosphere. All the other planets (and Ganymede) have magnetic fields of <2 G, requiring a space-based observatory for detection. For example, the magnetospheric radio emissions of Saturn, Neptune and Uranus were initially detected in-situ by Voyager during fly-bys (Zarka 1998).

These auroral processes are driven by a) magnetic reconnection between the planetary magnetic field and the magnetic field carried by the solar wind (e.g., Earth, Saturn, Neptune and Uranus), b) the departure from co-rotation with a plasma sheet residing in the planetary magnetosphere (e.g., Jovian main auroral oval) or c) interaction between the planetary magnetic field and orbiting moons (e.g., Jupiter-Io current system) (Zarka 1998 and references therein). As well as providing diagnostic information on the presence, strength, and extent of planetary magnetospheres, the detected radio bursts are the only means to accurately determine the rotation period of the planetary interior for the gas giants.

FARSIDE will be the first space-based radio telescope with sufficient sensitivity for detection of radio emission from all the solar system's magnetized planets from cis-lunar space. Monitoring the impact of the variable solar wind on the auroral radio flux density at the planets, including during interplanetary shocks from CMEs, is one application. Long-term monitoring of the Kronian radio emission period, modulated by the solar wind, will refine measurement of the true planetary rotation period (Zarka et al. 2007). The magnetospheric radio emissions from Neptune and Uranus have not been observed at radio wavelengths since Voyager.

In addition, Saturn and Uranus were found by Voyager to produce radio emission associated with atmospheric lightning (Zarka et al. 2004). This radio emission can potentially inform us about atmospheric dynamics and composition, when compared to optical imaging of cloud data (Zarka et al. 2012). The bursts are expected to have duration of 30 – 300 ms according to Zarka et al. (2004). The rms noise is 100 mJy in 60s at 200 kHz. As one example 300 ms bursts will be attenuated by a factor of 200x in a 60s image and thus the 1-sigma sensitivity to such bursts is 20 Jy. This is sufficient for frequent detection of SED and detection of the brightest UED bursts.

Finally, FARSIDE would offer the unique possibility of searching for radio emission from large bodies beyond Neptune out to hundreds of AU. This includes, but is not limited to, the putative Planet 9 (e.g., Batygin & Brown 2016). The expected flux density of Planet 9 is estimated by scaling from outer solar system planets assuming the radio emission is proportional to Poynting flux on the magnetospheric cross-section. A semi-major axis of 600 AU and radius of 3 Earth radii is assumed (Batygin, Brown & Becker 2019). Planet 9 can be interacting with the local ISM or the heliotail. For the heliotail, the B field and velocity are extrapolated from the inner solar wind. For the ISM, it is assumed that the magnetic field and density are consistent with measurements of the local ISM from Voyager 1 and 2 (Izmodenov and Alexashov 2020) and that the velocity of Planet 9 relative to the ISM is dominated by the motion of the Sun through the local ISM (~25 km/s). A magnetic to radio efficiency of $2 \times 10^{-3}$ is assumed in both cases (Zarka et al. 2018). A flux density of 10 mJy is estimated for the heliotail scenario and 3 Jy for the ISM interaction, if a significant planetary magnetosphere is present, although the uncertainty is very large.

### 4.2.3 The Magnetospheres and Space Weather Environments of Habitable Exoplanets

As described above, the radio emission from the Sun and planets provides key insights into the bulk motion of plasma in the solar corona and interplanetary medium and a means for direct measurement of planetary magnetospheres. Extending to other stellar and planetary systems in the solar neighborhood will provide essential data on the impact of energetic particles and coronal mass ejections on planetary atmospheres and habitability, together with the role of planetary magnetic fields as a buffer against such activity. The FARSIDE array, in particular, will be uniquely capable as an instrument for this science due to its ability to extend by a factor of a hundred times lower in frequency with equivalent or better sensitivity than current or planned ground-based arrays, enabling the routine detection of stellar CMEs, SEP events, and the magnetospheres of extrasolar terrestrial planets for the first time (Figure 12).

***Detection of Stellar CMEs and SEPs:*** As discussed in §4.2.1, solar CMEs and SEP events can be accompanied by radio bursts at low frequencies, particularly Type II bursts, as well as a subset of Type III bursts (complex Type III-L bursts). The emission is produced at the fundamental and first harmonic of the plasma frequency and provides a diagnostic of the density and velocity (few 100 to >1000 km/s) near the shock front, while the flux density of the burst depends sensitively on the properties of the shock and solar wind (Cairns et al. 2003). Ground-based radio astronomy can only trace such events to a heliocentric distance of a few solar radii, whereas space-based radio antennas can trace the propagation of shocks out to the Earth and beyond, which is particularly relevant for characterizing geoeffective CMEs and SEP events. The detection of equivalent interplanetary Type II and III events from stars other than the Sun is one of the goals of the FARSIDE array.

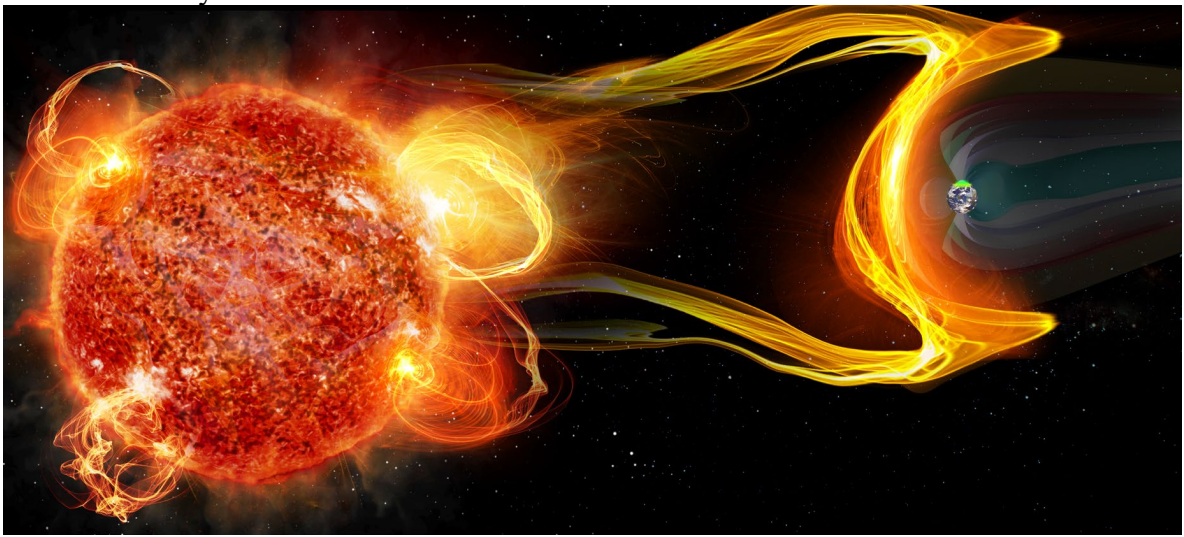

**Figure 12:** Artist's impression of a planet orbiting an M dwarf experiencing a CME due to a megaflare event. Credit: Chuck Carter and KISS/Caltech.

FARSIDE would detect the equivalent of the brightest Type II and Type III bursts out to 10 pc at frequencies below a few MHz. By imaging >10,000 deg$^2$ every 60 seconds, it would monitor a sample of solar-type stars simultaneously, searching for large CMEs and associated SEP events. For the $\alpha$Cen system, with two solar-type stars and a late M dwarf, it would probe down to the equivalent of $10^{-15}$ W m$^{-2}$ Hz$^{-1}$ at 1 AU, a luminosity at which solar radio bursts are frequently detected at the lowest frequencies available to FARSIDE (Krupar and Szabo 2018). The nearby young active solar-type star, $\varepsilon$Eridani (spectral type K2), is another priority target. For the case of M dwarfs, FARSIDE would be able to detect Type II bursts formed at the distance where super-Alfvénic shocks should be possible for M dwarfs (Villadsen & Hallinan 2019) and directly investigate whether the observed relationship observed between solar flares and CMEs (Aarnio et al 2011, Osten and Wolk 2015) extends to M dwarfs. If it does, FARSIDE should also detect a very high rate of radio bursts from this population.

***Detection of the Terrestrial Planet Magnetospheres:*** All the magnetized planets in our solar system, including Earth, produce bright coherent radio emission at low frequencies, predominantly originating in high magnetic latitudes and powered by auroral processes (§4.2.2; Zarka 1998). Detection of similar radio emission from candidate habitable planets is the only known method to directly detect the presence and strength of their magnetospheres. The temporal variability of this radio emission can also be used to determine the rotation periods and orbital inclinations of these planets. As described in §4.2.2, the radio emission is emitted at the electron cyclotron frequency (§4.2.2).

Extending to the exoplanet domain requires a very large collecting area at low frequencies, with the first detections likely to be ground-based. Indeed, in a recent breakthrough, radio emission has been detected from a possible free-floating planetary mass object (12.7 ±1 M$_{jup}$; Kao et al. 2018). This is the first detection of its kind and confirmed a magnetic field much higher than expected, >200 times stronger than Jupiter's, reinforcing the need for empirical data. However, the detection of the magnetic fields of candidate habitable planets will almost certainly require a space-based array if the magnetic fields are within an order of magnitude in strength of Earth's magnetic field. Detection of the magnetic field of planets orbiting in the habitable zone of M dwarfs is particularly key, as such planets may require a significantly stronger magnetosphere than Earth to sustain an atmosphere (Lammer et al. 2007).

The detection of radio emission from planets orbiting nearby stars is very sensitive to the stellar wind conditions imposed by its parent star and can serve as an indirect diagnostic of its velocity and density. It is possible to estimate the expected radio power from such planets, based on scaling laws known to apply to radio emission from the solar system planets (Zarka 2007 and references therein). These scaling relations are not only descriptive but predictive, with the luminosity of both Uranus and Neptune predicted before the Voyager 2 encounters and found to be in excellent agreement with the measurements (Desch and Kaiser 1984). Extrapolations to terrestrial exoplanets in the habitable zone of M dwarfs, and thus embedded within a dense stellar wind environment, predict radio luminosities that are orders of magnitude higher than the Earth (Farrell et al., 1999; Zarka et al. 2001; Grießmeier et al., 2007; Grießmeier 2015, Vidotto et al. 2019). During enhanced solar wind conditions, the Earth's radio emission can increase in luminosity by orders of magnitude (Figure 13) and a similar effect is expected for exoplanets. Therefore, as is the case for detecting Type II bursts associated with CMEs, it is essential to have the capability to monitor large numbers of planets simultaneously to detect periods where exoplanets greatly increase in luminosity.

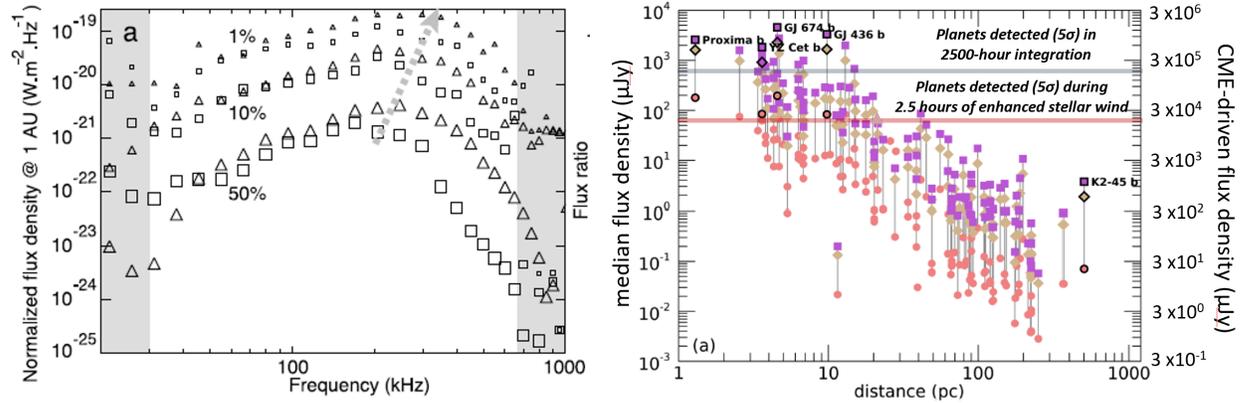

**Figure 13.** *Left*: The radio spectrum of Earth's auroral kilometric radiation for 50%, 10% and 1% of the time. The flux density across the entire detectable band can vary by factors of a few hundred (Lamy et al. 2010). *Right*: The predicted average radio flux density at ~280 kHz of known exoplanets orbiting M dwarfs from Vidotto et al. (2019) assuming a magnetic field of 0.1 G (10% of Earth's magnetic field). Purple, yellow, and red data points reflect different models for the quiescent stellar wind. The horizontal grey line represents the 5σ detection limit (y-axis on left-hand side) on the median flux density for the auroral radio emission detectable from FARSIDE in a 2500-hour integration. The horizontal red line represents the 5σ detection limit (y-axis on right-hand side) in a 2.5-hour integration during enhanced stellar wind conditions, e.g., during a CME, assuming a similar degree of variability as observed for Earth's auroral kilometric radiation. A large sample of planets, including some nearby candidate habitable planets (such as Proxima b), are detectable.

As noted in §4.1, FARSIDE will image 10,000 deg$^2$ of sky visible from the lunar farside every 60 seconds, in 1400 channels spanning 100 kHz–40 MHz. The sensitivity at the lower end of the band, enabled by the unique environment of the lunar farside, is particularly well suited to the search for radio exoplanets. In each FARSIDE image, there will be ~2000 stellar/planetary systems within 25 pc. At these low frequencies, the spatial resolution is limited to a few degrees due to scattering in the interstellar medium and interplanetary medium. However, with 1 stellar system within 25 pc per ~5 square degrees, identification of the source of detected emission should be straightforward, particularly as FARSIDE is most likely to detect exoplanets within 10 pc (Figure 13). An advantage at these low frequencies, is that the galaxy is optically thick at distances <100 pc, and thus long integrations are possible without classical confusion noise. Over the course of a two-year observing program, FARSIDE will have accumulated an average of 4000 hours of observing time on each of 8000 stellar planetary systems within 25 pc. Radio exoplanets will be identified by two means:

1. Deep 2500-hour integrations will be used to place limits on the average radio power from radio exoplanets.

2. Individual 2.5-hour integrations will be searched for heightened emission, expected during CME interaction.

Figure 13 highlights that FARSIDE would detect the radio emission from a population of Earth-sized, and super-Earth sized planets orbiting nearby M dwarfs, including a number of candidate habitable planets, providing the first measurements of terrestrial planet magnetospheres outside our solar system. The radio emission can be distinguished from that of the star by rotational modulation, as well as circular polarization which is largely absent for interplanetary solar radio bursts (Zarka et al. 2007; Hallinan et al. 2013). Detection of magnetospheres, if present, would identify the most promising targets for follow-up searches for biosignatures, as well as providing a framework for comparative analysis of exoplanet magnetospheres and atmospheric composition.

### 4.2.4 Sounding of the Lunar Subsurface

ROLSES will attempt to probe the subsurface layers beneath its lander. FARSIDE has the potential to sound the mega-regolith and its transition to bedrock expected at ~2 km below the surface (Hartmann 1973). The Lunar Radar Sounder (LRS) onboard the KAGUYA (SELENE) spacecraft has provided sounding observations of the lunar highlands (Ono et al. 2010) and found potential scatterers in hundreds of meters below the subsurface. However, the results are inconclusive due to surface roughness. FARSIDE, by virtue of being on the surface, would not be affected by roughness. Data from a calibration beacon in orbit could be synthesized to identify deep scatterers and the transition to bedrock at km depths by virtue of the low frequencies, which are significantly more penetrating. Deep subsurface sounding can also be performed passively using Jovian bursts from 150 kHz to 20 MHz (e.g., Romero-Wolf et al. 2015). The array covers a 10 km × 10 km area on the lunar highlands which could provide a three-dimensional image of the highland subsurface structure.

### 4.3 PRIME – Prototype Radio Interferometer on the Moon for Exoplanet studies

A pathfinder concept for FARSIDE, named PRIME, has been developed to perform initial low frequency radio science on the lunar surface and to reduce technology risks. PRIME would be the first operating radio interferometer on the Moon. This six-element radio array would operate over a frequency bandwidth of 0.1-20 MHz and a maximum baseline of 200-m. Use of a small rover will demonstrate the deployment of receiver/antenna nodes on the lunar surface. The rover will use a tether connected to the lander for power, communications, and data transmission. Thin wire, electrically-short, 10-m antennas will be embedded within the tether, following the design concept developed for FARSIDE. Investigation of the electrical properties of the antennas placed directly on the dielectric regolith will characterize the impact of the reflection of radio waves from the subsurface. The electronics used in the antenna nodes will have heritage from NASA's heliophysics mission SunRISE (Kasper et al. 2019), a six cubesat interferometer in high Earth orbit to measure compact structure in CMEs. Finally, PRIME will test the thermal designs and baseline concept of operations for FARSIDE. The technology objective is to validate the instrument concept by demonstrating background noise-limited performance.

PRIME will have sufficient sensitivity to produce dynamic spectra for transient emission from the Earth (if emplaced on the nearside), the Sun, and Jupiter. Of particular interest are Types II and III solar bursts. If launched in the next few years, PRIME will observe near solar maximum, during which ~3 CMEs leave the Sun each day, and will detect >100 Type II/III bursts, based on well understood power-law flux distributions. With Earth-Moon Very Long Baseline Interferometry (VLBI) between SunRISE and PRIME, this VLBI baseline would provide very precise distance estimates via curvature of the radiation wave front for complex Type III events, which will resolve a long-standing controversy concerning the origin of the electrons responsible for these events and their location relative to the CME shock front. PRIME would also characterize low frequency Jovian emission with a peak flux density of $10^7$ Jy (Carr et al. 1983), and if located on the near side, Earth's Auroral Kilometric Radiation (Alexander et al. 1975). Characterizing these emissions will serve to refine for exoplanet observations (Zarka 2007) with FARSIDE.

Taking a point source with a flux density of $10^9$ Jy as a representative test case, one can form a model PRIME beam at a given frequency, done here for 10 MHz. A notional PRIME layout is seen in Figure 14 on the left – one 200-m length arm with three pairs of antennas placed in the zig-zag pattern for orthogonal polarizations. The black dot represents the lander, and the six red stars are the nodes for the 10-m antennas, deployed along the tether shown by the black lines. For imaging, the software pipeline described in Section 4.1.3 was used along with the antenna locations and elevations shown in Figure 14. Then, the pipeline aligns reference frames and tracks the relative positions of the Earth, Moon, and Sun, and finally makes dirty images of the given input distribution. This reconstructed "snapshot" image is on the right.

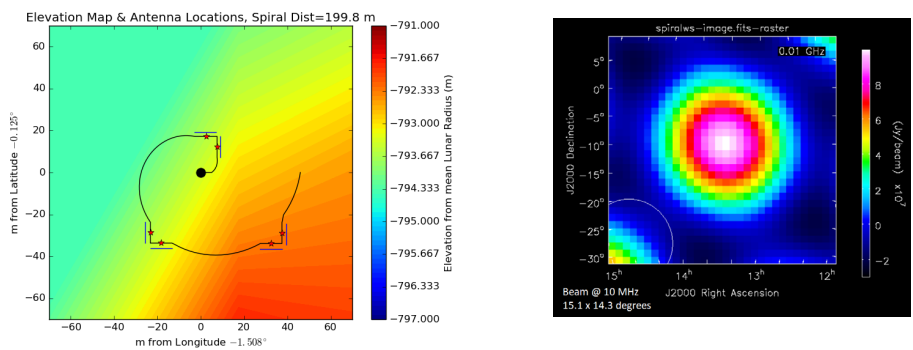

**Figure 14.** With the distribution of three pairs of PRIME antennas in a 200-m spiral pattern on the lunar surface shown in the left panel, the resulting interferometric "dirty" image from a "snapshot" at 10 MHz on the right panel illustrates the clear detection of a $10^9$ Jy point source as expected for a Type II solar radio burst. The white ellipse is the FWHM of the array beam.

## 5. Summary and Conclusions

After decades of study and numerous proposed concepts, radio astronomy on the Moon is finally underway. China has placed the first low radio frequency antenna on the lunar farside. This will soon be joined by NASA radio science instruments ROLSES and LuSEE. In this paper, we presented a roadmap for low radio frequency (∼0.1-40 MHz) observations from the Moon beginning with ROLSES and LuSEE, progressing to a six-element prototype radio interferometer (PRIME), and finally a 128× 2 antenna node array (FARSIDE) as a Probe-class mission to the lunar farside late in this decade.

ROLSES will fly first on the Intuitive Machines CLPS lander in late 2021. ROLSES will deploy two dipole antennas on the Moon's nearside, at 1-m and 3-m above the lunar surface. Its science goals include measurements of electrons and charged dust within a photoelectron sheath near the surface. These observations will provide key data to differentiate between a conflicting set of theoretical models. ROLSES will also measure the temporal and spectral characteristics of the current human-generated RFI from Earth to determine what solar and solar system observations will be possible from the lunar surface.

LuSEE is expected to launch in Q4 of 2024 and land within the Schrödinger impact basin on the farside of the Moon. Its suite of electromagnetic instruments will survey the lunar environment and determine if high dynamic range spectral and imaging observations are possible on the farside, in preparation for low radio frequency arrays. Like ROLSES, LuSEE will focus on the plasma environment, but on the farside surface. Sensitive measurements of the time-variability of the ionosphere will be made. In addition, dynamic spectra of radio bursts from the Sun and Jupiter, and the radio-quiet sky will be collected to set the stage for the PRIME and FARSIDE radio arrays.

PRIME is a six-node array of antennas that is proposed to fly on a CLPS lander to buy-down risks for the larger imaging array FARSIDE. PRIME will focus on proving the technologies for FARSIDE, including deployment of antennas, receiver boxes, and a tether, providing power and communications, using a teleoperated rover. It will demonstrate the expected noise-limited performance on the lunar surface. PRIME will observe Type II/III bursts from the Sun and radio emission from Jupiter to prepare for similar FARSIDE observations of low frequency radio radiation from CMEs in nearby stellar systems and auroral radio emission from exoplanets. FARSIDE will image the sky observable from the lunar farside every minute in 1400 spectral channels to search for signatures of bursting radio emission from several thousand nearby stars and exoplanets.

The approved PRISM CLPS missions described here, ROLSES and LuSEE, along with proposals for low radio frequency arrays, PRIME and FARSIDE, come at a time of increasing interest in lunar exploration and the Artemis human missions to the Moon. These signal a new paradigm for surface science and exploration from NASA, international partners, and commercial companies. Radio astronomy projects for the Moon take advantage of new developments in transportation, communications (e.g., Lunar Gateway), lunar operations, and infrastructure (e.g., Artemis). With these initiatives taking root and blossoming in this decade, astrophysical radio telescopes on the Moon come at an opportune time with an emerging architecture to achieve high-priority science objectives. With the development of new infrastructure at the Moon that may impact high sensitivity, long integration observations, it is important to build these telescopes soon and to develop international agreements on RFI mitigation.

## Acknowledgments

This work is directly supported by the NASA Solar System Exploration Virtual Institute cooperative agreement 80ARC017M0006. Support was also provided by the Simons Foundation grant "Planetary Context of Habitability and Exobiology". We are indebted to a number of highly-talented individuals who contributed to the mission and concept studies described in this paper. For ROLSES, this includes Damon Bradley, Pietro Sparacino, William Farrell, Richard Katz, Igor Kleyner, David McGlone, Michael Choi, Scott Murphy, and Rick Mills at NASA/GSFC. For LuSEE, this includes Peter Harvey, Keith Goetz, Lindsey Hayes, and Misty Willer. For FARSIDE and PRIME, this includes Larry Teitelbaum, James Lux, Andres Romero-Wolf, Issa Nesnas, Patrick McGary, and Tzu-Ching Chang at Caltech/JPL, as well as Steve Squyres and Alex Miller at Blue Origin Inc. We also thank the referees for their thoughtful comments which improved the manuscript.